\documentstyle[prl,aps,epsf,preprint]{revtex}

\begin{document}
\title{On the impossibility of temperature extraction from heavy ion
induced particle spectra}
\author{Jens Konopka, Horst St\"ocker, and Walter Greiner}
\address{Institut f\"ur Theoretische Physik,
         Postfach 11 19 32,
         Johann Wolfgang Goethe-Universit\"at,
         D--60054 Frankfurt am Main, Germany}
\date{\today}
\maketitle
\begin{abstract}%
Spectra of various particle species have been calculated with
the Quantum Molecular Dynamics (QMD) model for very central collisions
of Au+Au. They are compatible with the idea of a fully stopped thermal source
which exhibits a transversal expansion besides the
thermal distribution of an ideal gas. However, the microscopic
analyses of the local flow velocities
and temperatures indicate much lower temperatures at densities associated
with the freeze-out. The results express the overall impossibility of a
model-independent determination of nuclear temperatures from heavy ion
spectral data, also at other energies (e.g.\ CERN) or for other species
(i.e.\ pions, kaons, hyperons)!
\end{abstract}

The study of nuclear matter under extreme conditions is the basic motivation
for heavy ion experiments \cite{Sto86,Clare86}.
The aim is the creation of a large bulk of
heated and compressed nuclear matter \cite{NASI93}.
However, the temperature and the density are not straightforwardly
determinated from
these experiments. It is even not clear from the beginning, whether
it makes sense to speak about a temperature, since the degree of
equilibration in a single event is not a measurable quantity either.
In this letter
the relation between spectral information, which is measurable in
experiments, and the density and the degree of random motion during the
intermediate reaction stages is investigated using transport theory.
The latter aspect is closely related to the question of
thermalization and the possible analogy to thermostatic properties
of heated and/or compressed matter.

Single particle and $N$-body transport models have been used
extensively to gain insight into the stiffness of the equation of state,
the strength of the momentum dependence and in-medium nucleon-nucleon
scattering cross-sections \cite{Aic88,Pei89,Gale87,Blatt93,Faess93}.
These calculations predict various observables in terms of different
underlying assumptions. The question is, which of these aspects is
dominating the
dynamics of heavy ion collisions. This procedure is in fact very
indirect. It would be more convenient to have a direct measure of
thermodynamical quantities, like pressure or temperature, which could
be connected to the properties of excited nuclear matter.

As an idealized test case, we performed QMD calculations
\cite{Aic88,Pei89,Pei92}
of a heavy system for b=0, namely Au (150 MeV/nucleon) + Au.
Such reactions are
supposed to yield the largest accessible participant region, where the
nucleons involved undergo a rapid sequence of binary collisions.
Consequently, equilibrium, with a large number of nucleons involved,
is more likely established in central collisions than in
peripheral reactions or collisions of lighter partners.

In the present analysis no information other than mass, charge and
four-momentum of the emitted products will be used for extraction of further
information on the properties of the hot and dense nuclear matter formed in
the transient state. Thus such an analysis would be applicable to
experimental data as well \cite{Cof94,Lisa95}.

After a full QMD propagation for 300 fm/c, involving all nuclear interactions
and hard binary scatterings, fragments are calculated via a 3 fm
configuration space coalescence.
Then, in addition, all charged products are
propagated on their mutual Coulomb-trajectories for another $10^6$ fm/c
$\approx 3\cdot 10^{-18}$ s.

Having obtained the full triple-differential
spectra in such a manner for each charge independently, the spectra were
fitted by a transversally expanding, thermally equilibrated source
\cite{Kon94}.

This implies that the transverse spectrum is composed of a directed flow and
a random component, which can be associated with a temperature. The strength
of the collective component is assumed to increase linearly with the
transverse distance; the shape of the source is a homogeneous sphere.
Nonrelativistically the double-differential distribution reads
\begin{eqnarray}
\frac{{\rm d}^2 f(T,p_{\rm max})}{{\rm d}p_{\rm l}\ {\rm d}p_{\rm t}}
              = N \frac{2}{p_{\rm max}}\int_0^{p_{\rm max}} {\rm d}p_0 \;
              \frac{p_0}{p_{\rm max}}
              \left(\frac{\eta}{\pi}\right)^{\frac{3}{2}} 2\pi\ p_{\rm t}
              \nonumber \\
              \times \exp\{-\eta(p_{\rm t}^2+p_{\rm l}^2+p_0^2)\}\
              I_0(2\eta p_{\rm t}p_0)\
\end{eqnarray}
Here $\eta$ is an abbreviation for $\frac{A}{2 m_N T}$. With these formulas
the average flow momentum per nucleon is $\frac{2}{3}p_{\rm max}$ and the
averaged flow kinetic energy amounts to $p_{\rm max}/4m_N$.

In order to get information on the thermal collective energy sharing,
we restrict our analysis for the moment to the transverse momentum spectra
only. This procedure is motivated by the fact that transverse momenta
are newly created and are not directly connected to the initial (longitudinal)
beam momentum. A possible collective expansion in longitudinal direction
cannot unambiguously be associated to the properties of the hot and dense
reaction zone because the system may have memory about its history,
in particular the incident momentum.

Fig.\ \ref{gsipt} shows transverse momentum spectra of various fragment
charges obtained with QMD for the system Au (150 MeV/nucleon, b=0) + Au
(symbols) together with fits to these calculated data, which are based on the
assumptions from above. In fact, the corresponding count rates have been
fitted, rather than the invariant distributions, which are displayed.
All spectra are compatible with temperatures
between 20 and 25 MeV and averaged collective flow velocities of 0.1--0.13
c. Fig.\ \ref{papchi2} shows the minimum of the $\chi^2$
distributions associated with the various fits, which exhibit
a valley over a wide range of
temperatures. For heavy fragments such as $\alpha$'s or even heavier species,
the minimum is well defined. It is represented in the by the bullet:
For this class of probes the
collective component can be extracted for this model in an unambiguous way:
In the valleys of the $\chi^2$ distributions the corresponding flow momentum
depends only weakly on the temperatures. On the contrary, for light
particles like free protons and deuterons one observes that
\\
1) the minimum is not as pronounced as in the case of the heavy
composite fragments, particularly the fit for p's is of considerable minor
quality and is not unique
\\
2) the valley extends for p's
like a banana from low temperatures with large flow
to high temperatures with almost no flow with only a slight gradient.
This implies that the heavy fragments clearly exhibit collective flow in these
ultracentral collisions, while the flow cannot
unambiguously be determined with light fragments.
Collective flow is here associated with azimuthally symmetric
collective expansion.

Fig.\ \ref{papchi2} shows the line corresponding to energy conservation
When the fit procedure works well, the
minimum of the $\chi^2$ distribution is in line with energy conservation.
However, this is not the case for protons, which is a first indication that the
proton emission pattern cannot be understood in terms of the simple
ansatz of collective flow + thermal motion, although the proton spectrum in
Fig.\ \ref{gsipt} seems to be properly fitted.

Fig.\ \ref{papchi2} indicates that the heavier
fragments exhibit smaller averaged transverse collective expansion velocities:
In the model these fragments form via the configuration space
coalescence mechanism in the interior of the expanding nuclear matter. On the
surface of the exploding system, where the local expansion velocity is highest,
it is less probable to form heavy fragments, since the number density of
nucleons around is much smaller than in the interior.

Let us summarize the analysis of the final spectra, apparently the
momentum space population of all fragment species is reasonably
well described by two parameters, the temperature
$T$ and an averaged collective velocity
$v_c$. The spectral information suggests temperatures larger than
20 MeV. Since this depends only on the final state of the collision, when
all interactions on the nuclear length scale have ceased, it is concluded
that the ashes of the exploding nuclear matter seem to indicate these
high temperatures.
The associated average collective velocity amounts to 10-15\% of the speed of
light. Quantitatively similar conclusions have been obtained from measured
transverse
momentum and global kinetic energy spectra in central collisions of Au+Au
at 150 MeV/nucleon\cite{Cof94,Jeo93,Rei94}.
However, a detailed analyses of the nuclear chemistry,
i.e.\ the fragment abundances, in the very same
experiment yields low
temperatures of the order of 8 MeV to account for the large
fragment abundances\cite{Kuh93}.
This has lead to speculations on different time scales for the chemical
and the dynamical equilibration processes.
The temperatures determined with the two methods, chemical determination
from isotope ratios or fragment abundances on the one hand side and
determination from the fragment spectral slopes on the other hand can not
be reconciled in this scenario. We demonstrate below that this is due to
misconceptions of the interpretation of yields and spectra in an
instantaneous break-up of a homogeneous, globally equilibrated system.

QMD allows for a microscopic calculation of all quantities
relevant to the reaction dynamics and the nuclear thermodynamics. The time
evolution of the composite system is followed in more detail \cite{Kon94} by
dividing configuration space in various cells on a cylindrical grid.
For each of these cells the mean and
the spread of the {\em local} velocity distribution
is calculated for a superposition of many
events, obtained under the same macroscopic conditions.
The mean velocity illustrates
the collective velocity of each individual cell, whereas the spread around
the mean reveals the thermal, i.e.\ random,
motion inside the cell. The directional dependences
of the temperature parameters (longitudinal, transversal, and azimuthal
temperatures $T_{\parallel},~ T_{\perp},~ T_{\varphi}$) have also been
extracted. They correspond to the widths of the local velocity distribution
in the directions parallel to the beam, perpendicular to the beam,
and in azimuthal direction. These values are equal in each cell
in the case of local equilibrium. If the temperatures
were, in addition, constant over the full volume, this would be
global thermal equilibrium. The complete information on the local collective
velocities, and on the degree of randomness is available. It
tests our initial assumption of a linearly increasing flow velocity
profile and we can analyze to which extend global or local thermal
equilibrium is established in the model.

In the following we characterize the central reaction zone in
terms of the global parameters density and temperature, which are
important for the (multi-)fragmentation of the source.
The central reaction zone is defined here as the volume
of all cells with density at least half of the maximum density at this
instant. This definition implies a time-dependent volume,
which is considered here.

A considerable fraction of the nucleons remains in the central volume
(Fig.\ \ref{tevol} a)): even after 100 fm/c about one third of the system
is still in the central zone.
Matter outside this zone has experienced
the higher densities earlier on. In terms of a hydrodynamical
point of view, matter in the QMD model freezes out at different times.
The temperature distributions indicate rapid cooling and
prove that the source of nucleons and clusters is
not globally equilibrated, even for the most central collisions
in microscopic models (Fig.\ \ref{tevol} b)).
However, in the longitudinal direction the spread of the velocity
distribution is higher by 50\% in the intermediate reaction
stages (t=40--60 fm/c), while the system is transversally equilibrated.
This cooling is associated with a decrease of the density too
(Fig.\ \ref{tevol} c)).

Fig.\ \ref{paprho-t} shows that at low densities $(<0.5\rho_0)$, where the
fragment freeze-out is commonly assumed to occur, the
{\em microscopically} determined true (local) temperatures (symbols)
have dropped below 10 MeV. Thus they strongly deviate from
the slope temperatures extracted from the simple {\em macroscopic} fit to
the (microscopically calculated) fragment spectra (light grey shaded area).
The values compare reasonably well with
curves of constant entropy obtained with the quantum statistical model
(QSM) \cite{Hahn88} and with a QSM analysis of
experimental fragment yields \cite{Kuh93} (dark shaded area).

This explanation of
the large discrepancy between the temperatures extracted from the
chemical analysis and the temperatures extracted from the fragment spectra
puts into question all temperature extractions from slopes
of particle spectra, even if an additional collective motion is
taken into account. This was done, for example, in our macroscopic analysis
above. The values are 2--3 times higher than the correctly extracted
microscopic values. This finding concerns also temperatures extracted
from slopes of pion spectra at highest incident energies, e.g.\
at the AGS and CERN \cite{Hei92}.

What is wrong with the assumptions
underlying the global fit?
A thermally equilibrated
source at some temperature $T$
can disintegrate not only due to the thermal pressure, but also due to
some additional collective expansion. The probability
of finding a particle which has received a collective velocity
$v_{\rm coll.}$, i.e.\ $dN/dv_{\rm coll.}$, is essentially unknown.
It can be expressed as
\begin{equation}
\frac{{\rm d}N}{{\rm d}v_{\rm coll.}} = \frac{{\rm d}N}{{\rm d}r} \cdot
\frac{{\rm d}r}{{\rm d}v_{\rm coll.}}
= \frac{{\rm d}N}{{\rm d}r} \cdot \left(\frac{{\rm d}v_{\rm coll.}}{{\rm d}r}
\right)^{-1}\, .
\end{equation}
Hence, the density distribution as well as the flow velocity
profile enters as the first and second term on the right hand side of the
preceding equation. For a fit to the spectra, a homogeneous density with
a sharp cut-off and a linearly increasing velocity profile have been used
\cite{Sto81a}.

Therefore, the high transverse
momenta components of the particle spectra do not correspond to the
high momentum tails of hot source, but to the collective motion of
fast cells with low internal temperature.
The microscopic analysis suggests that the combinations of
density and temperatures, which are traversed in the course of the reaction,
are in agreement with the expectations from the quantum statistical analysis
of the fragment distributions in the final state \cite{Kuh93}.

In summary, temperatures obtained from slopes of particle spectra
can only serve as an upper estimate for the conditions at freeze-out,
even if collective flow has been taken into account.
The main origin of this uncertainty is the fact that the
configuration space distribution is essentially unknown.
In view of this analysis, model estimates of temperatures in
high-energy and ultrarelativistic heavy ion collisions \cite{Hei92}
are expected to overestimate the thermal energy reached
in such reactions considerably.
Our analysis suggests that chemical thermometers, e.g.\ $\pi/p$ ratios, which
take into account feeding from unstable states and absorption may be more
appropriate.

\section*{Acknowledgement}
This work was supported by BMFT, DFG, and GSI.

\newpage
\begin{figure}
\caption{Invariant transverse momentum distributions of various charged
fragments emerging from Au (150 MeV/nucleon, b=0) + Au collisions. The
lines indicate fits to the calculated points involving a azimuthally
symmetric, transversally expanding, and thermally equilibrated source.
The fits have been performed for each fragment specie independently.
The associated count rates rather than the displayed invariant
distributions have been fitted. All spectra are compatible with a
temperature between 20 and 25 MeV and averaged collective flow
velocities between 0.10 and 0.13 c.}
\label{gsipt}
\end{figure}

\begin{figure}
\caption{$\chi^2$ distributions for the two parameter fit to the spectra
in Fig.\ \protect\ref{gsipt}. Contourlines are at 2, 3, 4, and 5
times the minimum value. The minima are indicated as black circles. The
thick solid lines correspond to those pairs of parameters where the total
kinetic energy is conserved. For the case of protons, the quality of the
fit in terms of $\chi^2$ is poor. The energy constraint tends to predict
lower temperature and larger flow.}
\label{papchi2}
\end{figure}

\begin{figure}
\caption{Thermodynamics in the central reaction zone, i.e.\ the volume
where the local density is at least half of the maximum value. Mass
content a), spreads of the local velocity distributions b), and maximum
as well as averaged density c) are displayed at a function of time.}
\label{tevol}
\end{figure}

\begin{figure}
\caption{Temperatures versus densities for central Au+Au collisions
at 150 MeV/nucleon. The microscopically determined local temperatures
(symbols as in Fig.\ \protect\ref{tevol}) are considerably lower than
the simple macroscopic fit to the spectra (light grey shaded area).
However, they compare reasonably well with curves of constant entropy
from QSM and with experimental data from a quantum statistical analysis of
the fragment distributions (dark grey shaded area) \protect\cite{Kuh93}.}
\label{paprho-t}
\end{figure}

\newcommand{\figref}[1]{\newpage
                        \begin{center}
                        \mbox{\epsffile{#1.eps}}
                        \par\vskip 2cm
                        \large\bf J.\ Konopka et al., Fig.\ \ref{#1}
                        \end{center}}

\figref{gsipt}
\figref{papchi2}
\figref{tevol}
\figref{paprho-t}

\end{document}